\documentclass[conference]{IEEEtran}
\usepackage{amsmath,amssymb,amsfonts}
\usepackage{algorithmic}
\usepackage{graphicx}
\usepackage{textcomp}
\usepackage{multirow}
\usepackage{tabularx}
\usepackage{booktabs}
\usepackage{amsmath}
\usepackage{amsthm}
\usepackage{color}
\usepackage{enumitem}

\usepackage{caption}
\usepackage{subcaption} 
\usepackage{listings}
\usepackage{url}
\usepackage{pgfplots}
\usepackage{color}
\usepackage{cite}
\DeclareCaptionFont{black}{\color{black}}
\DeclareCaptionFormat{listing}{\colorbox{gray!50}{\parbox[l]{\columnwidth}{#1#2#3}}}
\usepackage[]{algorithm2e}
\usepackage{algorithmic}

\renewcommand{\baselinestretch}{0.98}

\lstdefinelanguage{solidity}{
    alsodigit = {-},
    keywords = {function,uint,internal,pure,returns,string,if,return,while,bytes,memory,new,return,byte,struct,address,uint256,bool,assert}
}

\lstdefinelanguage{graphviz}{
	alsodigit = {-},
	keywords = {Node,->,lable,True Branch,False Branch}
}

\def\BibTeX{{\rm B\kern-.05em{\sc i\kern-.025em b}\kern-.08emT\kern-.1667em\lower.7ex\hbox{E}\kern-.125emX}}
    
\DeclareCaptionFont{black}{\color{black}}
\DeclareCaptionFormat{listing}{\colorbox{gray!50}{\parbox[l]{\columnwidth}{#1#2#3}}}
\usepackage[]{algorithm2e}
\usepackage{algorithmic}

\begin{document}

%
\title{SIF: A Framework for Solidity Contract Instrumentation and Analysis}

%

\author{\IEEEauthorblockN{Chao Peng}
\IEEEauthorblockA{\textit{University of Edinburgh} \\
Edinburgh, United Kingdom \\
chao.peng@ed.ac.uk}
\and
\IEEEauthorblockN{Sefa Akca}
\IEEEauthorblockA{\textit{University of Edinburgh} \\
	Edinburgh, United Kingdom \\
	s.akca@sms.ed.ac.uk}
\and
\IEEEauthorblockN{Ajitha Rajan}
\IEEEauthorblockA{\textit{University of Edinburgh} \\
Edinburgh, United Kingdom \\
arajan@ed.ac.uk}
}

%

%

%

%
\maketitle

\begin{abstract}
Solidity is an object-oriented and high-level language for writing smart contracts that are used to execute, verify and enforce credible transactions on permissionless blockchains. 
In the last few years,  analysis of smart contracts has raised considerable interest and numerous techniques have been proposed to check the presence of vulnerabilities in them. 
Current techniques lack traceability in source code and have widely differing work flows. There is no single unifying framework for 
analysis, instrumentation, optimisation and code generation of Solidity contracts. 

In this paper, we present SIF, a comprehensive framework for Solidity contract analysis, query, instrumentation, and code generation. SIF provides support for Solidity contract developers and testers to build source level techniques for analysis, understanding, diagnostics, optimisations and code generation. We show feasibility and applicability of the framework by building practical tools on top of it and running them on 1838 real smart contracts deployed on the Ethereum network. 
 

\end{abstract}

\begin{IEEEkeywords}
high level languages, software testing, code instrumentation, program analysis
\end{IEEEkeywords}

\section{Introduction}
\label{sec:introduction}
Blockchains are the underlying technology for making online secure transactions using cryptocurrencies such as Bitcoins and Ethers. 
Executing, verifying and enforcing credible transactions on blockchains is done using smart contracts, which is code written by the buyer and seller using Turing-complete languages~\cite{buterin2014next}. 
Solidity is a popular object-oriented and high-level language for writing smart contracts~\cite{dannen2017introducing, tsankov2018securify} and can be compiled to bytecode for execution on the blockchain network. 
With the increased use of smart contracts across application domains, there is a crucial need for a unified framework that supports and facilitates Solidity code analysis, understanding, transformation, and development of tools for verification and testing that provide strong security guarantees. 

\subsection{Problem}
\label{sec:problem}
Traditional programming languages are generally supported by a comprehensive framework for code instrumentation, monitoring, optimisation and code generation such as LLVM/Clang~\cite{LLVM:CGO04} for C/C++. This framework support is lacking for smart contract programming languages like Solidity. 
Existing work and tools for Solidity contracts is primarily based on analysing the bytecode of smart contracts for security vulnerabilities, or translating Solidity code to other languages or intermediate forms over which security analysis is performed~\cite{luu2016making, tsankov2018securify, mythril, remixdocumentation,bhargavan2016formal, tikhomirov2018smartcheck, kalra2018zeus}.
For instance, Oyente~\cite{luu2016making} analyses bytecode from the Solidity compiler and successfully reports the presence of vulnerabilities. 
However, it lacks the ability to trace and localise bugs in Solidity code.
Zeus~\cite{kalra2018zeus} translates smart contracts to LLVM bitcode but does not support complete Solidity syntax, including \texttt{throw} statements, \texttt{self-destructs}, virtual functions and  assembly code blocks.
These techniques, although effective in detecting security vulnerabilities, are not general purpose code instrumentation and analysis tools. As a result, building program analysis tools with a purpose different from detecting security vulnerabilities is difficult using existing techniques. Traceability back to Solidity source code is also a common issue in these techniques. 
We propose a general purpose instrumentation and analysis framework that addresses this need for Solidity contracts.


\subsection{Goal}
\label{sec:contribution}

We seek to provide Solidity developers and testers with a framework, similar to the widely-used C/C++ instrumentation technique Clang LibTooling~\cite{libtooling}, to easily and effectively understand, manipulate and analyse Solidity code.
To achieve this goal, we design and implement \emph{Solidity Instrumentation Framework (SIF)} with the following capabilities:
\begin{enumerate}
  \item Provide an interface for users to query the Abstact Syntax Tree (AST) of Solidity code. 
  \item Support code instrumentation or transformation using pre-defined helper functions that can modify the AST. 
  \item Support Solidity code generation from AST. 
\end{enumerate}


\noindent\textit{Tools.}
We demonstrate the general purpose nature of the SIF framework in supporting code instrumentation and analysis by using it to implement 7 different tools or utilities for Solidity contracts. Four of these tools query the Solidity abstract syntax tree (AST) to provide information on the contract code and three others instrument and modify the code using the AST. A brief description of the tools is provided below,  
\begin{description}
 \item[Function Listing] lists all the function definitions with function names, return lists, parameter lists and which contract they belong to. 
Useful in summarising and reviewing Solidity contracts.
 \item[Call Graph Generator] produces function call graphs depicting calling relationships between functions in a Solidity contract. Nodes represent functions and edges represent the calls relation. Useful in understanding and reviewing smart contracts. 
 \item[Control Flow Graph Generator] produces a graphical representation of control flow within a Solidity contract. Control flow graphs are very useful in static analysis and program optimisations. 
 \item[AST Diff] is a syntactic diff tool for comparing Solifity contracts at the AST level. It ignores comments, white spaces and blank lines. 
 \item[SIF Rename] allows the user to easily rename existing identifiers. All definitions and references to the specified identifier will be changed to the new name provided.
 \item[Fault Seeder] allows developer to artificially seed bugs in smart contracts. The tool currently supports 3 common bug/vunerability types. The seeded bugs can be used to assess effectiveness of testing or verification tools in uncovering vulnerabilties.  
 \item[Assertion Analyser] examines all AST nodes of the smart contract and determines if the node is susceptible to vulnerabilities, such as division by zero, overflow and underflow. 
Assertions will be inserted by the tool at contract locations susceptible to these vulnerabilties. Assertions will be checked during execution of the modified contract. 
\end{description}

We evaluate SIF and the 7 tools built on top of it on a collection of 1838 real smart contracts that are running on the Ethereum network.
Our results show that SIF and the utility tools are easy to use, highly automated and new tools can be easily implemented with the helper functions.

\section{Related Work}
\label{sec:related}
Several static analysis and vulnerability detection tools for smart contracts have emerged in the last few years. We present and discuss existing work over  
Solidity contracts for vulnerability detection, code generation, query and instrumentation.

\paragraph{Vulnerability detection}
Many of the existing techniques rely on  bytecode analysis to check for potential vulnerabilities~\cite{luu2016making, tsankov2018securify, mythril, remixdocumentation}. 
Bytecode is the compiled hexadecimal format of smart contracts.
Solidity contract verification at the bytecode level is sufficient to detect vulnerabilities, However, it is challenging to trace the problem back to Solidity source code, that the developer can then fix. 
Some techniques translate Solidity code to the F* languages or represent code in an intermediate form, such as  LLVM IR (Intermediate Representation) or XML~\cite{bhargavan2016formal, tikhomirov2018smartcheck, kalra2018zeus}. Vulnerability analysis is then done over the intermediate representation.
The techniques do not support translation of the intermediate form back to Solidity, which results in loss of traceability in source code. 

\paragraph{Solidity code generation from AST/Intermediate form}
Currently, translation from intermediate form or AST back to Solidity has limited support.
We are only aware of one tool, Soltar~\cite{soltar}, that supports translation of Solidity AST back to Solidity code. 
However, Soltar does not provide capabilities supported by SIF to query and modify the AST.
Additionally, Soltar is not maintained and does not support Solidity versions 0.4.3 onwards. 
SIF, presented in Section~\ref{sec:overview}, is able to handle the latest Solidity version 0.5.3, and older versions.

An existing tool, Zeus~\cite{kalra2018zeus}, supports translation of Solidity code to LLVM bitcode but does not have the capability to generate Solidity back from bitcode. Additionally, complete Solidity syntax is not supported (\texttt{throw} statements, \texttt{self-destructs}, virtual functions and  assembly code blocks) and there is no provision for code instrumentation. 
Porosity~\cite{porosity} and an Ethereum smart contract decompiler, JEB~\cite{jebdecompiler}, translate bytecode back to Solidity code. However, support for Solidity code query, modification and instrumentation is not available in these tools.

\paragraph{Query and instrumentation of Solidity code}
There exists no general purpose framework or tool that provides capabilities for users to query, modify and instrument Solidity code. These capabilities are crucial for program understanding, analysis, optimisation and verification techniques~\cite{ryder1979constructing,allen1970control,geimer2009generic,tikir2002efficient}.  The SIF framework is novel in providing these capabilties and in providing interfaces that allow users to easily define the kind of query, modification and instrumentation needed.  
SIF's capabilities have allowed us to query and instrument Solidity code to produce call graphs, control flow graphs, function listing, Fault Seeder, Assertion Analyser along with scope for building many custom utilities and diagnotics. 
We discuss SIF's design and the tools built on it in the next Section. 


\section{SIF Overview}
\label{sec:overview}
\begin{figure*}
	\centering
	\includegraphics[trim = 1cm 0cm 1cm 0cm, scale=0.5]{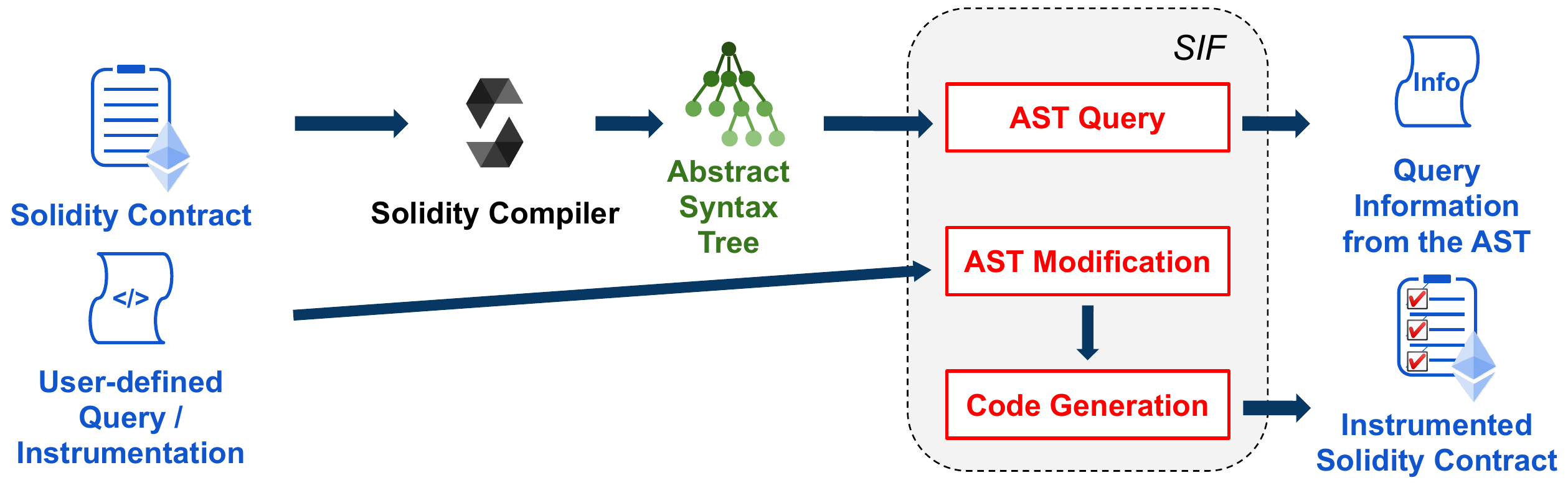}
	\caption{SIF work flow}
	\label{fig:sif-work-flow}
  	\vspace{-15px}
\end{figure*}

In this section, we present our generic framework, SIF, implemented in C++. SIF enables Solidity code query and instrumentation and works at the AST level. The framework also provides the capability for generating Solidity code back from the AST. SIF supports the entire Solidity syntax, up to version 0.5.3.
The workflow of SIF is illustrated in Figure~\ref{fig:sif-work-flow}.
SIF starts from the AST of Solidity code, produced by the Solidity compiler. It then accepts user instructions on queries and/or modifications needed. The framework then gathers the desired query information or performs modifications to the AST and finally generates Solidity code from the AST.

In the rest of this Section, we describe the design SIF, how it can be used for query and instrumentation along with instructions needed from the user, a concrete illustration of the query and instrumentation capabilities using seven tools that we built on top of SIF. The framework with its source code and user guide is available at \url{https://github.com/chao-peng/SIF}. To allow users to try the framework without having to download and build the source code, we have provided an online version of SIF at \url{https://wandbox.org/permlink/PnaL6bO9zipKRuKu}.

\subsection{SIF Design}
\label{sec:implementation}

Operations of SIF are dividied into 3 phases. Phase 1 focuses on representing AST nodes as C++ classes with methods to retrieve and modify information of the node. Phase 2 interacts with user defined query and/or instrumentation functions, and traverses the AST to perform the desired operation. Phase 3 generates Solidity code from the AST. We discuss each of these phases in more detail in the rest of this Section. We use an example \texttt{struct definition} named \texttt{Request} shown in Listing~\ref{lst:struct-example}, containing a data element and a method, to illustrate the different phases. 
%
%

\begin{lstlisting}[language=solidity,caption={Struct example},label={lst:struct-example}]
struct Request {
  bytes data;
  function (bytes memory) external callback;
}
\end{lstlisting}

\begin{figure}
\centering
   \includegraphics[trim = 1cm 0cm 1cm 0cm, scale=0.4]{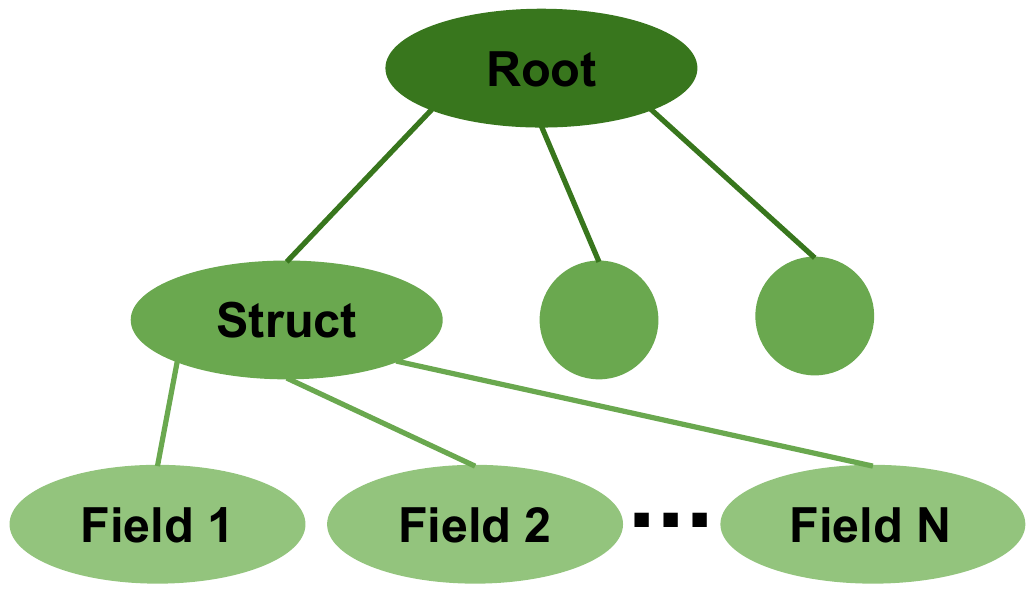}
   \caption{An AST example highlighting a struct definition}
  \label{fig:AST}
  	\vspace{-15px}
\end{figure}

\noindent\textbf{\textit{Phase 1: AST Representation.}}
The Solidity compiler generates the AST from Solidity code in two formats: plain text and JSON (JavaScript Object Notation, a structured data format).
In an attempt to make our tool for code instrumentation generic and easy to use, we use C++ classes as our intermediate AST representation. 
Given the Solidity AST, SIF first traverses the AST  and for each node, it instantiates a class of that node type with all the associated information. Figure~\ref{fig:AST} shows a struct definition appearing in an example AST. 
The classes contain information about the AST node in their data fields, and provide methods to query and modify the data.
Listing~\ref{lst:struct-representation} shows the C++ class representation of a struct definition node in the AST. The C++ class provides methods for getting name of the struct, setting it to a different name, querying number of fields in the struct, getting a particular field, adding, removing and updating a field. 
These methods facilitate query and modification of the AST node. 

\begin{lstlisting}[language=c++,caption={Representing structs using a C++ class},label={lst:struct-representation}]
class StructDefinitionNode : public ASTNode {
public:
  StructDefinitionNode() : ASTNode(); // Constructor
  string source_code(); // produce source code 
  string get_name(); // get the name
  void set_name(new_name); // update the name
  int num_field(); // get the number of fields
  void add_field(new_field); // add a new field
  void remove_field(index); // remove a field
  void update_field(index, _new); // update a field
  ASTNode get_field(index); // get a field by
private:
  string name; // name of the struct
  vector<ASTNode> fields; // list of struct fields
};
\end{lstlisting}
%
%

\noindent\textbf{\textit{Phase 2: Query and Instrumentation.}}
SIF interacts with users for AST query and instrumentation through a function \textit{visit} that it declares. 
SIF traverses the AST starting from the root node in a depth-first fashion using the \textit{visit} function. The user can implement queries and modifications, if any, that are needed within the \textit{visit} function. 
For example, the user can change the name of the example struct \textit{Request} from Listing~\ref{lst:struct-example} to \textit{DirectRequest} by implementing it in the \textit{visit} function, as shown  in Listing~\ref{lst:struct-rename}. Each time SIF visits an AST node, the \textit{visit} function is called to process operations defined by the user.
In this example, the \textit{visit} function first determines whether the current node is a \texttt{struct definition} and then checks whether the struct name matches \textit{Request}. 
If there is a match, the struct name  is changed to \textit{DirectRequest}. The methods, \textit{get\_name() and set\_name()}, used in this implementation are helper methods provided by SIF to facilitate writing of AST queries and instrumentations.

\begin{lstlisting}[language=c++,caption={Change the name of a struct in SIF},label={lst:struct-rename}]
void visit(ASTNode* node) {
  if (node->get_node_type() == StructDefinition) {
    StructDefinition* sd = (StructDefinition*) node;
    if (sd.get_name() == "Request") 
      sd.set_name("DirectRequest");
  }
}
\end{lstlisting}
%

\noindent\textbf{\textit{Phase 3: Solidity code generation.}}
For each type of AST node classes, SIF defines code templates that allow corresponding Solidity code to be generated. 
Listing~\ref{lst:source-code} illustrates the code template for the \texttt{StructDefinitionNode}. 
In the listing below, the template generates Solidity code for a \texttt{struct definition} by first printing the \texttt{struct} keyword, followed by the name of the struct, a left brace indicating the start of struct element definitions.
The template then iterates through the list of struct element definitions. Once an element definition is visited, the source code of that definition is appended to the struct source code. Once all the element definitions are completed,  a right closing brace is added, indicating the end of the struct definition.
If the name of the struct were changed by the user, as shown in Listing~\ref{lst:struct-rename}, the new struct name will appear in the generated Solidity code.
\begin{lstlisting}[language=c++,caption={Source code template of structs},label={lst:source-code}]
string StructDefinitionNode::source_code() {
  string source = "struct " + name + "{\n";
  Iterate the list of element definitions:
    source += element.source_code() + ";\n";
  return source + "}\n";
}
\end{lstlisting}

\subsection{Using SIF}
\label{sec:components}

SIF provides an interface in the form of predefined functions - \textit{before}, \textit{visit} and \textit{after}, to perform user-defined AST queries and instrumentations. We have presented in Section~\ref{sec:implementation} how the \textit{visit} function can be used for this purpose.
%
The \textit{before} function is called by SIF in advance of traversing the AST, and can be used for data initialisation. The \textit{after} function is called when visiting the AST is completed and helps summarise information gathered from the AST.
In the following Section, we demonstrate how these three functions can be used in custom utility tools built with SIF. 

\subsection{Tools using SIF}
\label{sec:extra-tools}
\begin{figure}
	\centering
	\includegraphics[trim = 1cm 0cm 1cm 0cm, scale=0.55]{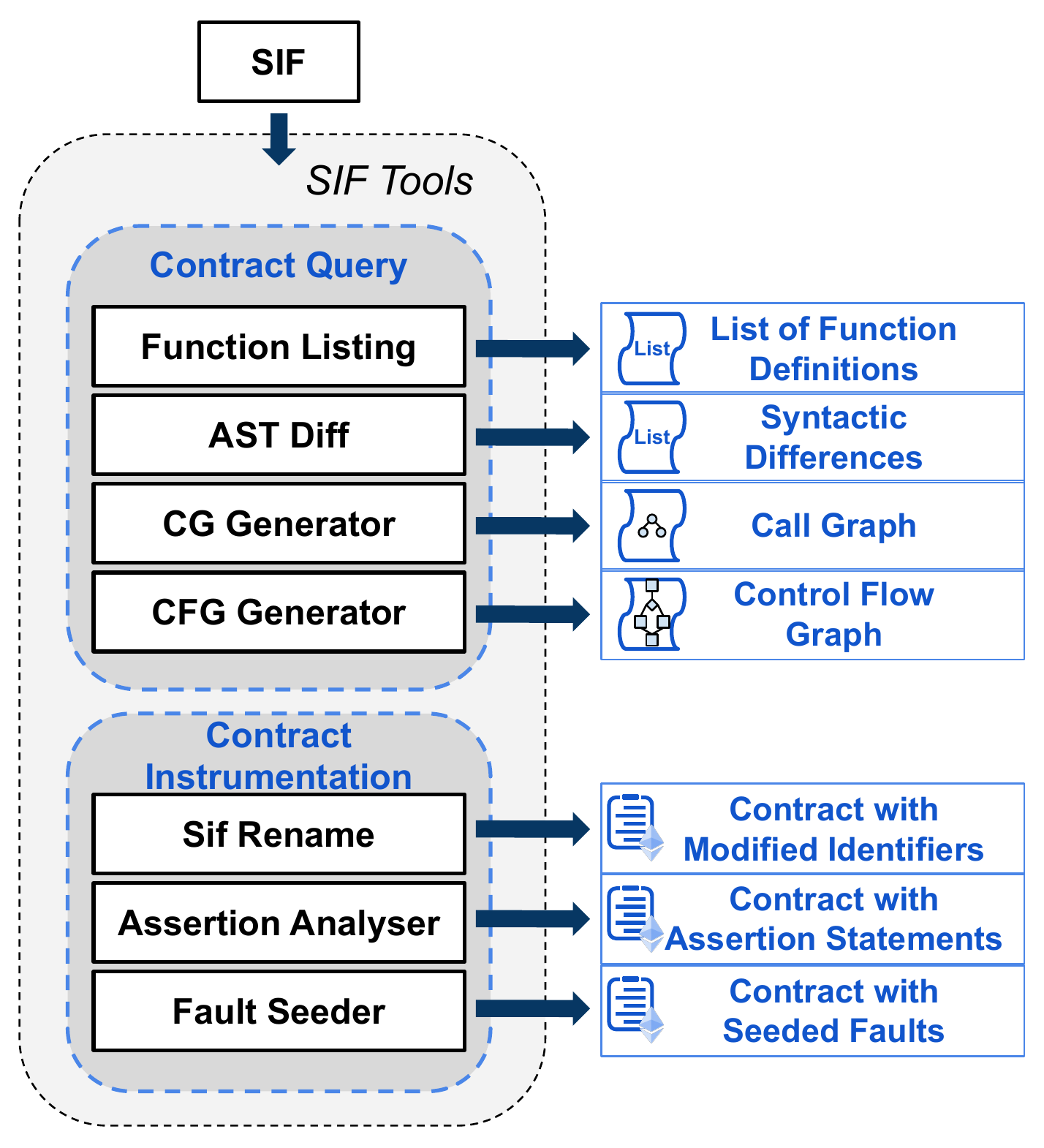}
			  	\vspace{-5px}
	\caption{Using SIF to build tools}
	\label{fig:tools-work-flow}
	  	\vspace{-15px}
\end{figure}

To illustrate the framework's extent of support for building custom queries and instrumentations, we implemented seven utility tools for Solidity contracts, as shown in Figure~\ref{fig:tools-work-flow}. The seven tools were inspired by tools built over other mature code instrumentation frameworks, such as Clang LibTooling for C/C++. We discuss and present each of these tools in the rest of this Section. 
%
\paragraph{Function Listing}
Given a Solidity file with multiple contracts and function definitions, this tool outputs a list of functions that appears in the file with function names, parameter lists, return lists and the contracts in which they are defined. 
It is worth noting that in Solidity, functions are considered members of contracts and appear as children of contract definition nodes in the AST representation.
To implement the \texttt{Function Listing} tool, we use the pre-defined \textit{visit} function within SIF. We check whether the visited node is a contract definition node. For contract definition nodes, we check all the children nodes to see if they are of type function definition. If function definition nodes exist, we record information on its name, parameters, and return values using SIF's helper methods for this node type. Once SIF finishes traversing the AST, the recorded information on functions definitions is printed. 
%
Listing~\ref{lst:function-listing-example} shows the list of function definitions printed by this tool for the AztraToken contract in Etherscan\footnote{The contract AztraToken is available at \url{https://etherscan.io/address/0x6962E259a8f9633C4494764628A7984cCEd58e10}}.

\begin{lstlisting}[language=solidity,caption={Function definitions summerised by Function Listing},label={lst:function-listing-example}]
[In AztraToken] AztraToken() returns ()
[In AztraToken] _transfer(address _from, address _to, uint _value) returns ()
[In AztraToken] transfer(address _to, uint256 _value) returns ()
[In AztraToken] transferFrom(address _from, address _to, uint256 _value) returns (bool success)
[In AztraToken] approve(address _spender, uint256 _value) returns (bool success)
[In AztraToken] burn(uint256 _value) returns (bool success)
[In AztraToken] burnFrom(address _from, uint256 _value) returns (bool success)
[In AztraToken] mintToken(address target, uint256 mintedAmount) returns ()
[In AztraToken] freezeAccount(address target, bool freeze) returns ()
[In AztraToken] transferOwnership(address newOwner) returns ()
\end{lstlisting}

\paragraph{CG Generator}

\begin{figure*}
	\centering
	\includegraphics[trim = 1cm 0cm 1cm 0cm, scale=0.45]{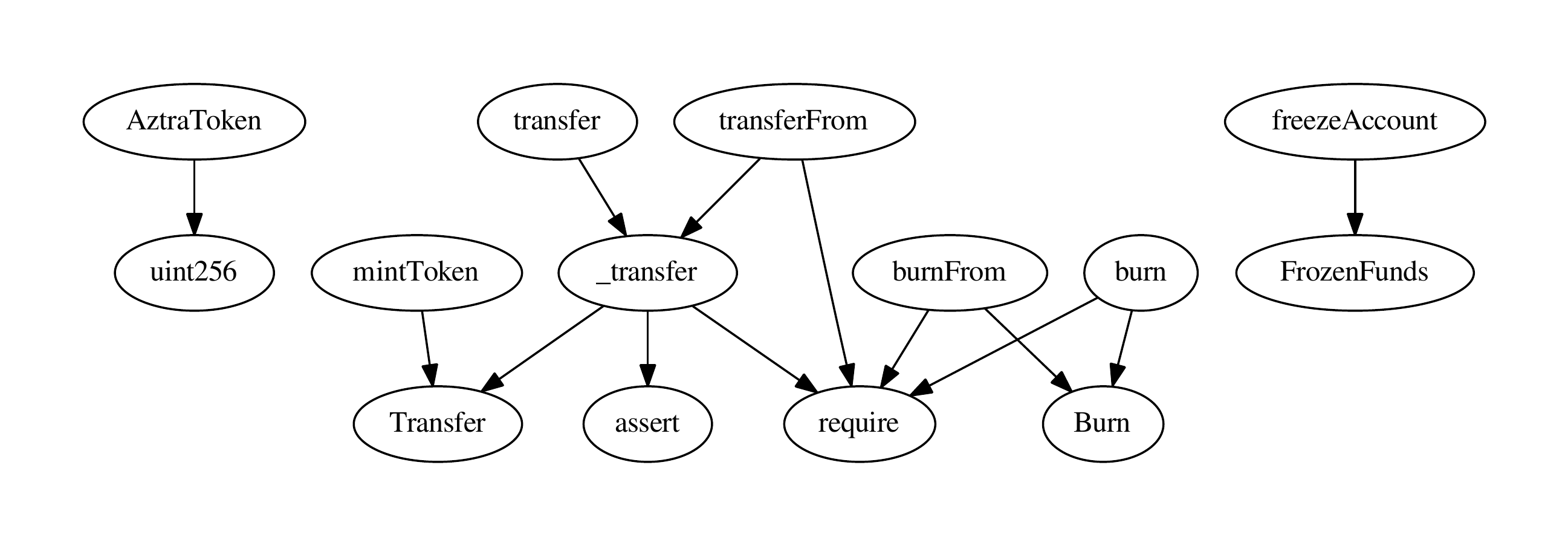}
			  	\vspace{-10px}
	\caption{Call Graph of the Smart Contract AztraToken}
	\label{fig:CG-example}
	  	\vspace{-15px}
\end{figure*}

\texttt{CG Generator} illustrates the calling relationships between functions as a call graph (CG). The tool is implemented using the \textit{visit} function first to process two types of AST nodes: function definition and function call. The tool maintains a map containing the callee function with the caller functions it is associated with. 
After traversing the full AST, the graph is drawn by the \textit{after} function using a graph drawing tool, Graphviz~\cite{ellson2001graphviz}, with function names as nodes and edges based on the relations recorded in the map.
Figure~\ref{fig:CG-example} presents the call graph generated by our tool for the AztraToken contract.

\paragraph{CFG Generator}

\begin{figure}
	\centering
	\includegraphics[trim = 3cm 0cm 3cm 0cm, scale=0.35]{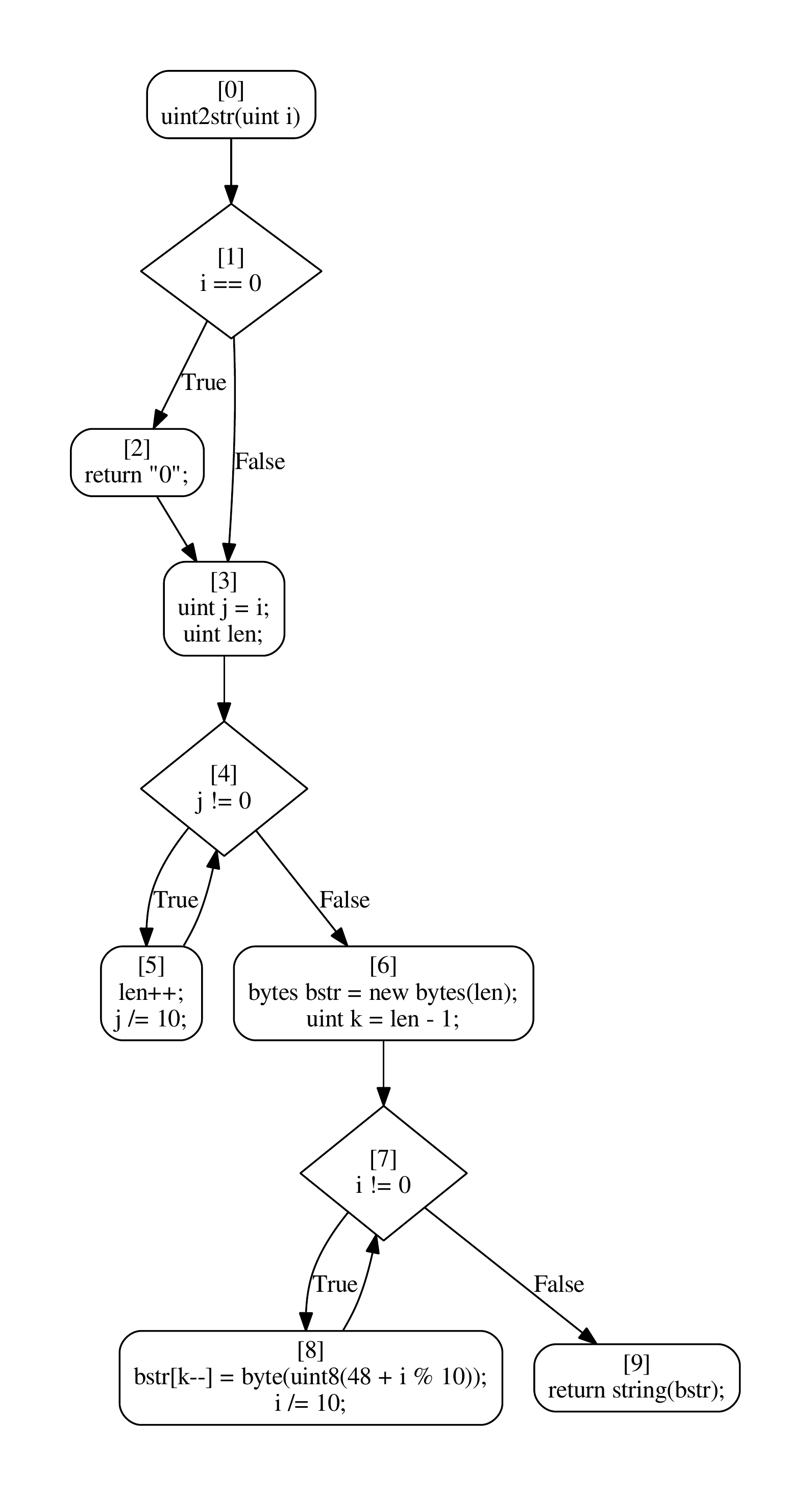}
		  	\vspace{-10px}
	\caption{Control Flow Graph of the Function uint2str from the Smart Contract Item}
	\label{fig:CFG-example}
	  	\vspace{-15px}
\end{figure}

Control flow graphs (CFG) are used to illustrate control flow within a program~\cite{allen1970control} and is fundamental to many static analysis and compiler optimisation techniques. 
The tool \texttt{CFG Generator} produces control flow graphs for functions or subroutines within Solidity contracts. Figure~\ref{fig:CFG-example} shows the CFG generated by our tool 
for the \textit{uint2str} function, shown in Listing~\ref{lst:example-control-flow}, that converts an unsigned integer variable to a string. 
The \textit{uint2str} function is found in contract \textit{Item} from Etherscan\footnote{The contract Item is available at \url{https://etherscan.io/address/0x5f896c654a08323dbe16aded331c461ccaeeb370}}.
\begin{lstlisting}[language=solidity,caption={Function to generate a control flow graph using CFG Generator},label={lst:example-control-flow}]
function uint2str(uint i) internal pure returns (string) {
  if (i == 0) return "0";
  uint j = i;
  uint len;
  while (j != 0){
    len++;
    j /= 10;
  }
  bytes memory bstr = new bytes(len);
  uint k = len - 1;
  while (i != 0){
    bstr[k--] = byte(48 + i % 10);
    i /= 10;
  }
  return string(bstr);
}
\end{lstlisting}

\texttt{CFG Generator} traverses the AST and clusters nodes in a straight line code sequence into basic blocks.
When a node introducing control flow is encountered, the most recent basic block ends and is linked to the condition node with the control flow.
The true branch of this control flow is linked to the first basic block of the then branch for an if statement or the loop body if it is a loop.
The false branch is linked to the first basic block of the else branch for an if statement or the first basic block immediately after the loop body if it is a loop.
Basic blocks and links between them are maintained and recorded in a data structure as the AST is traversed. Once the AST traversal is completed, the \textit{visit} function passes the data structure to Graphviz~\cite{ellson2001graphviz} to generate a graph depicting basic blocks and control flow. The data structure and resulting CFG is shown in the Listing~\ref{lst:example-control-flow-datastruct} and Figure~\ref{fig:CFG-example}, respectively. 
%
\begin{lstlisting}[language=graphviz, caption={Data structure produced by CFG Generator},label={lst:example-control-flow-datastruct}]]
Node [0] -> Node [1];
Node [1] -> Node [2] label=True Branch;
Node [1] -> Node [3] label=False Branch;
Node [2] -> Node [3];
Node [3] -> Node [4];
Node [4] -> Node [5] label=True Branch;
Node [4] -> Node [6] label=False Branch;
Node [5] -> Node [4];
Node [6] -> Node [7];
Node [7] -> Node [8] label=True Branch;
Node [7] -> Node [9] label=False Branch;
Node [8] -> Node [7];
\end{lstlisting}

\paragraph{AST Diff}

\texttt{AST Diff} takes two Solidity contracts and compares them at the AST level to report syntactic differences.
The tool ignores differences in comments, white spaces, and empty lines.
Implementation using the \textit{visit} function starts from the root node of one smart contract. Every node visited within this smart contract is compared with nodes in the other contract. Differences in data fields within the node are reported. Extra or missing nodes are also shown. 

\paragraph{SIF Rename}
This tool allows users to change the names of functions, variables and contracts by locating and replacing all occurences. 
The tool accepts three additional inputs: identifier type, old name and new name. The \textit{before} function first parses the command line options for these additional inputs.  As SIF traverses the AST, 
the \textit{visit} function checks if the currently visited node matches the type and name of the identifier provided by the user. If a match is found, the old name gets replaced with the new one.
The replacement is also performed within other nodes that reference the provided identifier.
For instance, if a user wants to rename a particular contract, \texttt{SIF Rename} will rename the contract within the \texttt{contract definition} AST node. It will also rename references to the contract within other nodes such as \texttt{member access} and \texttt{using for directive} that may reference the given contract.

\begin{table*}[th]
\centering
\begin{tabular}{|c|c|c|}
	\hline
	Type of Vulnerability                                                                   & Assertions Added by Assertion Analyser                                                                                                             & Fault seeded by Fault Seeder                                                                                                    \\ \hline
	Division by Zero                                                                        & require(c != 0); a = b / c;                                                                                                                        & Statement with division by zero is inserted.                                                                                    \\ \hline
	\multirow{2}{*}{Unsigned Overflow}                                                      & a  = b + c; assert(a \textgreater{}= b \&\& a \textgreater{}= c);                                                                                  & \multirow{2}{*}{\begin{tabular}[c]{@{}c@{}}Arithmetic operation resulting in \\ overflow is inserted.\end{tabular}}             \\ \cline{2-2}
	& \begin{tabular}[c]{@{}c@{}}a = b * c;\\ (b != 0 \&\& c != 0)? assert(a \textgreater{}=b \&\& a \textgreater{}= c): assert(a == 0);\end{tabular}   &                                                                                                                                 \\ \hline
	Unsigned Underflow                                                                      & a = b - c; assert(b \textgreater{}= a \&\& b \textgreater{}= c);                                                                                   & \begin{tabular}[c]{@{}c@{}}Arithmetic operation resulting in \\ underflow is inserted.\end{tabular}                             \\ \hline
	\multirow{3}{*}{\begin{tabular}[c]{@{}c@{}}Signed Overflow / \\ Underflow\end{tabular}} & \begin{tabular}[c]{@{}c@{}}a = b + c;\\ assert((c \textgreater{}= 0 \&\& a \textgreater{}= b) \textbar\textbar\space (c \textless\space 0 \&\& a \textless\space b));\end{tabular}   & \multirow{3}{*}{\begin{tabular}[c]{@{}c@{}}Arithmetic operation resulting in \\ overflow / underflow is inserted.\end{tabular}} \\ \cline{2-2}
	& \begin{tabular}[c]{@{}c@{}}a = b - c;\\ assert((c \textgreater{}= 0 \&\& a \textless{}= b) \textbar\textbar\space(c \textless\space 0 \&\& a \textgreater\space b));\end{tabular} &                                                                                                                                 \\ \cline{2-2}
	& \begin{tabular}[c]{@{}c@{}}a = b * c;\\ (b != 0 \&\& c != 0)? assert((a / b == c) \&\& (a / c == b)): assert(a == 0);\end{tabular}                    &                                                                                                                                 \\ \hline
\end{tabular}


 \caption{Vulnerability types}
 \label{tab:mutation}
 		  	\vspace{-15px}
\end{table*}

{\textbf{\textit{Fault Seeder and Assertion Analyser}}} These tools focus on 3 types of security vulnerabilities commonly found in smart contracts, presented in Table~\ref{tab:mutation}. 
Fault seeder instruments the code by injecting vulnerabilities. Assertion Analyser uses AST query and instrumentation capabilities to insert assertions in the contract. 

\paragraph{Fault Seeder}
For each type of vulnerability illustrated in Table~\ref{tab:mutation}, the \texttt{Fault Seeder} tool creates a new code block containing that vulnerability and injects it into the Solidity contract. 
SIF traverses the contract AST and inserts an extra AST node containing a single vulnerability into the original AST. 
Solidity code is then generated from the modified AST, and is referred to as a mutated contract. 
For example, to introduce integer underflow, \texttt{Fault Seeder} inserts the code snippet shown in Listing~\ref{lst:underflow-example} into the original smart contract. 

\begin{lstlisting}[language=solidity,caption={Code snippet to introduce underflow by Fault Seeder},label={lst:underflow-example}]
uint256 minuend = 20;
uint256 subtrahend = 250;
uint256 result = minuend - subtrahend;
\end{lstlisting}


\paragraph{Assertion Analyser}
Inserting property checks using assert statements is helpful in verifying program correctness. 
The \texttt{Assertion Analyser} tool inserts pre- and post-conditions for arithmetic operations to help detect overflow, underflow and division by zero vulnerabilities. 
Within the \textit{visit} function, the \texttt{Assertion Analyser} tool first gathers information on the node, and operators and operands if any within. 
For example, an expression node with arithmetic operations may be prone to an overflow/underflow error.  The tool then inserts relevant assert statements, shown in Table~\ref{tab:mutation}, as extra nodes following the node under analysis.
For division by zero vulnerability, SIF inserts a pre condition, before the node under analysis with a division operator, that asserts the divisor expression is greater than zero. 
Consider the function \textit{uint2str}, shown in Listing~\ref{lst:example-control-flow}, with a subtraction operation on line 10.  The \texttt{Assertion Analyser} tool will insert an assert statement, shown in Listing~\ref{lst:assert-example}, to check for unsigned integer underflow.  
\begin{lstlisting}[language=solidity,caption={Inserted guard for a subtraction operation by Assertion Analyser},label={lst:assert-example}]
uint k = len - 1;
assert(len >= k && len >= 1);
\end{lstlisting}

\section{Evaluation}
\label{sec:evaluation}
We evaluate feasibility, ease of use and extent of automation in using SIF and the seven tools built on it over 1838 unique Solidity contracts from the Ethereum network~\footnote{Our dataset is available at \url{https://github.com/chao-peng/SIF}}. 
%

Additionally, after SIF was released on Github in May 2019, we received various enquiries about using the framework from different parts of the globe. The interest expressed in the framework was encouraging. The feedback helped us fix bugs and improve the usability of the framework and is discussed in Section~\ref{sec:feedback}.

\subsection{Research Questions}

We investigate the following questions in our experiment:
\begin{description}[topsep = 0pt, itemsep = 0pt]
	\item[Q1. Solidity Language Support:] 
	\textit{Does SIF support all constructs in the Solidity syntax and generate code accordingly? }
	To answer this question, we first inspect contract code in our dataset and the official Solidity documentation to check the range of syntactic structures present in the dataset. 
	We then use SIF, without any instrumentations, to generate Solidity contracts from ASTs and we compare if the generated code is the same as the original.
	
	\item[Q2. Correctness of AST Query and Instrumentation:] 
	\textit{Are the tools built on top of SIF able to query and instrument the AST and produce correct output? }
	We use a small user group to help answer this question. The user group comprised of ourselves (\texttt{User-0}), 2 PhD students familiar with Solidity language (\texttt{User-1} and \texttt{User-2}), and a second year undergraduate student (\texttt{User-3}) with basic knowledge of Solidity. 
	We manually check the correctness of the query and instrumentation information over the different contracts. 
%
	
	\item[Q3. Feasibility, Extent of Automation and Ease of Use:] 
	\textit{How feasible is the framework for users to write their own tools? Are the tools fully automated and easy to use? }
	We asked \texttt{User-1}, \texttt{User-2} and \texttt{User-3}  to evaluate the framework and the 7 tools on the following aspects:
	\begin{enumerate}
		\item \textbf{Ease of implementing new tools} On a scale of 1 (very hard) to 5 (very easy), rate how easy it is to implement new tools using SIF.
		We asked the users to write a new tool, \texttt{Loop Count}, to report the number of loops in a Solidity contract, that utilises the AST query capability of the framework. 
		We also asked the users to create a tool, \texttt{Make Signed}, that changes all unsigned integer (uint) types to signed integer (int), utilising both query and instrumentation capabilities. 
		\item \textbf{Extent of automation of the 7 tools} On a scale of 1 (Completely manual) to 5 (fully automated), rate the extent to which the seven tools can be run automatically. 
		\item \textbf{Ease of use of the 7 tools} On the scale of 1 (Very difficult to use) to 5 (Very easy to use), rate the ease with which the 7 tools can be used.
	\end{enumerate}
	%
%
\end{description}


\subsection{Experimental Result}
\label{sec:result}
We discuss the experimental results in the context of the experiment questions presented earlier in this Section.

\subsubsection{Solidity language support}

The 1838 Solidity contracts in our dataset contain a wide variety of Solidity constructs. 
We manually checked against the Solidity documentation as well as the source code of the official Solidity compiler, and confirmed that our dataset covers all syntactic elements in Solidity.
Our technique was able to analyse the AST and generate source code for all 1838 smart contracts from their ASTs automatically. 
The generated contracts were also compared by the 3 users to the original contracts, and no differences were reported. 
As mentioned in Section~\ref{sec:related}, the F* tool~\cite{bhargavan2016formal} does not support loop structures in Solidity.
A significant fraction of contracts in our dataset contain loops and SIF is able to fully support loops.
Self-destructs, throw statements and inline-assembly blocks are not supported by the Zues verification tool~\cite{kalra2018zeus}. 
Our framework fully supports the use of these constructs in Solidity contracts. 

\subsubsection{Correctness of AST query and instrumentation}

The seven tools discussed in Section~\ref{sec:extra-tools} utilise SIF capabilities for  AST query and instrumentation. We evaluate correctness of these capabilities in the context of the seven tools. 

To validate output of tools \texttt{Function Listing} and \texttt{CG Generator}, we split the dataset into two halves. 
With contracts in the first half of the dataset, two users ran the tools to collect function lists and call graphs. The other two users validated the output produced by the tools by inspecting the original contract.  
We, then,  swapped the roles of the users for contracts in the second half. 
This was done to reduce biases, if any, and the monotony in checking the contracts. 
The users did not find any anomalies in the tools over the given dataset. 

Control flow graphs produced by \texttt{CFG Generator} were manually inspected by \texttt{User-0} to check its correctness. CFG generator correctly generated CFGs for all contracts in the dataset. 
To evaluate \texttt{AST Diff}, we split the dataset into two halves again.  \texttt{User-0} and \texttt{User-3} modified contracts in the first half by renaming identifiers, mutating operators and deleting code blocks for the first half. \texttt{User-1} and \texttt{User-2} ran the \texttt{AST Diff} on each set of original and modified contract (from first half), to check if all the modifications were reported correctly by the tool. The user roles were swapped for contracts in the second half of the dataset.  
AST Diff was found to work correctly by all users over all the contracts. 

For the instrumentation capability, we used \texttt{SIF Rename} to locate and change names of identifiers within contracts in the dataset.
We again, split the user reponsibilities so that two users selected and changed identifier names and two users checked the changes were reflected in the generated Solidity code. 
Different identifier types were selected. The users found SIF Rename worked correctly for the given dataset and requested changes. 

Each of the four users ran \texttt{Fault Seeder}, selecting different vulnerability types to be seeded in each Solidity contract, and inspected the mutated contracts to check if the vulnerabilities were seeded as per choice. 
The mutated contracts were also run using the Solidity compiler, \texttt{solc}, to check for syntactic correctness. No issues were reported. 
The \texttt{Assertion Analyser} tool was run on all contracts, and the users individually inspected if the assertions were inserted correctly for the arithmetic operations. The instrumented contracts with assertions were checked through the \texttt{solc} compiler for syntactic correctness; no problems were found.  The users were able to confirm the assertions were inserted by the tool correctly around arithmetic operations in all contracts.




\subsubsection{Extent of automation and ease of use}

\begin{table}[th]
	\centering
	\begin{tabular}{|c|c|c|c|c|}
		\hline
		\multirow{3}{*}{User \#} & \multicolumn{2}{c|}{Existing 7 Tools}                                                                                                                                  & \multicolumn{2}{c|}{Using SIF}                        \\ \cline{2-5} 
		& \multirow{2}{*}{\begin{tabular}[c]{@{}c@{}}Extent of \\ Automation\end{tabular}} & \multirow{2}{*}{\begin{tabular}[c]{@{}c@{}}Ease of \\ Use\end{tabular}} & \multicolumn{2}{c|}{Ease  of Implementing New Tools} \\ \cline{4-5} 
		&                                                                                  &                                                                         & Loop Count                & Make Signed               \\ \hline
		1                   & 4                                                                              & 4                                                                     & 4                       & 4                       \\ \hline
		2                   & 5                                                                              & 4                                                                     & 5                         & 4                         \\ \hline
		3                   & 4                                                                                & 5                                                                      & 4                         & 3                         \\ \hline
		Avg.                   & 4.3                                                                              & 4.3                                                                     & 4.3                         & 3.7                         \\ \hline
	\end{tabular}
	\caption{User experience of using SIF and tools}
	\label{tab:user-rate}
			  	\vspace{-5px}
\end{table}

Table~\ref{tab:user-rate} reports how \texttt{User-1}, \texttt{User-2}, and \texttt{User-3} rate the existing 7 tools for ease of use and extent of automation. The table also shows how the users rate feasibility and ease with which new tools, \texttt{Loop Count} and \texttt{Make Signed}, can be implemented using SIF.
As the user interface for the existing 7 tools is similar, we report the average ratings across the 7 tools for ease of use and extent of automation for each user.
\texttt{User-1, User-2, User-3} rate the extent of automation for the 7 tools as 4, 5 and 4, respectively, giving a high average user experience of 4.3 (mostly fully automated). This implies that the users agree that the tools have a high level of automation with little user intervention. 
The users gave a score less than 5 as the users had to use the Solidity compiler to first produce the AST and then call the tool. We will look to automate this step in the future so the users do not have to explicitly call the Solidity compiler.  

For ease of use of the 7 tools, all three users felt the interface for running the tools was easy to understand and use (ratings of $4, 4$ and $5$). 

For writing new tools with SIF, all three users felt it was easy to implement the \texttt{Loop Count} tool (average implementation ease being 4.3) owing to implementation similarity to the existing \texttt{Function Listing} tool (replace functions by loops). The users found implementing the \texttt{Make Signed} tool was slightly more challenging, as it is different from implementations of the existing 7 tools. Implementing the \texttt{Make Signed} tool requires searching through variable declarations, function parameter lists  to locate all unsigned integer qualifiers and replacing them with the signed qualifier. \texttt{User-3} found implementation of the \texttt{visit} function was only slightly easy (rating of $3$), while \texttt{User-1} and \texttt{User-2} found it easy (rating of $4$). 
{User-3} is not proficient in C++, and found the use of pointers in the \textit{visit} function interface tricky.
To address this issue, we have added comments in the interface showing how to use and cast to AST node pointers. 
Average ease of implementation across the users was 3.7 for the \texttt{Make Signed} tool.


Finally, we report the time taken in using the framework to generate Solidity code from unmodified ASTs.
For contracts with less than 1000 lines of code (90.0\% of the dataset), the framework finished code generation in 4 seconds. One of the bigger contracts, 
Expiry\footnote{The contract Expiry is available at \url{https://etherscan.io/address/0x0ece224fbc24d40b446c6a94a142dc41fae76f2d}} with 6183 lines of code, takes 85.6 seconds for code generation from AST. To understand why code generation for this contract took significantly longer, we profiled the framework phases and found that 40.3 seconds was spent on parsing the JSON format AST.
Our framework uses the text format AST as the main reference and only queries the JSON format AST when it encounters keywords missing in the text AST. 
However, each such query starts from the root of the JSON AST for each newly visited node, resulting in significant overhead. Other contracts with more than 5K lines of code finished code generation in less than 15 seconds as they did not have to refer to the JSON AST very frequently. In our future work, we will optimise and accelerate queries involving JSON data structures.

\subsection{Feedback from the Global Community}
\label{sec:feedback}
In the 2 months since SIF was realeased on GitHub, we noted interest in its use and received queries from users around the world. In this Section, 
we present feedback and suggestions from global users.

\paragraph{Bug report related to unitinitialised variables} This was reported by a user from Ohio State University. 
We found the bug was caused by un-initialised smart contract variable definitions. The framework looks for the missing initialisation statement in such variable definitions. We have now fixed this bug. 

\paragraph{Enquiry on gathering information from the AST} 2 users from China did not know how to maintain and summarise information gathered from different AST nodes. We provided them with guidance and examples of using \textit{before} and \textit{after} functions to address this issue. We also updated our documentation to include this. 

\paragraph{Request for more Command Line options} A user from Hong Kong University of Science and Technology made this request. After each SIF run, smart contract source code is generated by default. However, code generation is not always desired, especially for users who only want to query information from the AST. We have now provided this option to avoid code generation. We plan to add more command line options to SIF for additional user control.

\paragraph{Request for detailed documentation} A user from ShanghaiTech University requested more detailed documentation on each type of AST node and the associated fields and methods. We have provided this documentation for all types of AST nodes, along with examples using them.

\section{Conclusion}
\label{sec:conclusion}
SIF provides the capability to query,  analyse and instrument the AST of Solidity contracts and generate Solidity code back. The framework uses a C++ intermediate representation, and provides helper methods to gather information on different AST nodes and manipulate them, as needed.
The framework is generic and eases the implementation of custom query and instrumentation capabilties. Users interact with the framework using a simple interface that is easy to understand and use. We built 7 tools for different types of Solidity code query and instrumentation to evaluate SIF's versatality and asked 3 users to rate the usability and automation of the tools along with ease of implementing new tools. We used 1838 unique contracts in our evaluation. 

We found SIF was able to run all 1838 contracts, and the outputs of all 7 tools were confirmed to be correct by all the users. Additionally, the users found the 7 tools were ease to use and nearly fully automatic. The users also implemented 2 new tools using SIF and found the framework was well suited for implementing new tools, owing to the helper functions and simple interface. SIF's generality and ease of use makes it a promising framework for Solidity developers and testers to quickly and easily build, maintain  new tools or leverage existing capabilities. Since the release of the framework on GitHub, we have received interest in its use for Solidity code analysis and instrumentation from several international users. We envision SIF becoming a large tooling infrastructure like LLVM/Clang, supporting diagnostics, easy development and maintenance of Solidity tools. 

\bibliographystyle{ieeetr}
\bibliography{references}

\end{document}